\documentclass[twocolumn,prb,showpacs]{revtex4}
\usepackage{graphicx}
\usepackage{amsmath}
\begin{document}
\title{Diffraction and boundary conditions in semi-classical
 open billiards}
\author{T. Blomquist}
\affiliation{Department of Physics (IFM),
Link\"{o}ping University, S--581\,83 Link\"{o}ping, Sweden}
\date{\today}
\begin{abstract}
The conductance through open quantum dots or quantum billiards shows
fluctuations, that can be explained as interference between waves
following different paths between the leads of the billiard. We examine
such systems by the use of a semi-classical Green's
functions. 
In this paper we examine how the choice of boundary
conditions at the lead mouths affect the diffraction. We derive a new
formula for the $S$-matrix element. Finally we compare semi-classical
simulations to quantum mechanical ones, and show that this new
formula yield superior results.
\end{abstract}
\maketitle
Lateral quantum dots, also called quantum billiards serve as model
systems in the study of the relation between quantum and classical
physics. A very large number of experiments have been done on transport
through open semiconductor quantum dot,\cite{Marcus,Persson,
Boggild,Andy_sq} and also experiments on analog
systems such as microwave billiards have been performed.\cite{Stoeckmann,
Kim,Barth,Hersch}
These studies  
show rapid conductance oscillations as function of energy or of an
applied magnetic field. To provide a quantitative description of these
oscillations, several approaches have been adopted, such as random
matrix theory,\cite{Beenakker} numerical solution of the Schr\"odinger
equation\cite{Persson,Berggren,Ensslin,PRL,AkisBird}  and
semi-classical(SC) methods. In the SC view, the conductance
oscillations arise due to interference between pairs of
classical trajectories, carrying a quantum mechanical(QM)
phase,\cite{Miller,Smilansky,Baranger} between the leads of the
billiard.\cite{Gutzwiller}  
The SC approach has been used to describe statistical properties of
the conductance oscillations\cite{Smilansky,Baranger} including
weak-localization line shapes in chaotic and 
regular cavities, and fractal conductance in systems with mixed
phase space. One should however be careful in relying on these
results.\cite{TB2}  
The SC approach can also provide an interpretation of specific
frequencies in the conductance oscillations, by relating them to
specific classical trajectories in a billiard.  
In calculating conductance or transmission amplitudes of a system, a
SC approximation of the systems Green's function is used. 
In relating this Green's function to the transmission amplitude
between leads of the billiard, one has to take into account
diffraction effects at the lead mouths. 
In this paper we study how the choice of boundary conditions(BC) on
the Green's function affect the diffraction and make comparison to QM
calculations. 

We study a billiard with hard walls and a zero inner potential. The
Hamiltonian inside the billiard is
\begin{equation}
H=\frac{(-i\hbar\nabla+e\mathbf{A})^2}{2m^*},
\label{Hamilton}
\end{equation}
where $\mathbf{A}$ is the vector potential of a magnetic field, $e$ is
the electron charge and $m^*$ is the electron's effective mass.
The dynamics is decided by the Schr\"odinger equation
\begin{equation}
(H-E)\psi=0.
\label{Schrodinger}
\end{equation}
We define the Green's function
\begin{equation}
(H-E)G(\mathbf{q},\mathbf{q'})=\delta(\mathbf{q},\mathbf{q'}).
\label{Green}
\end{equation}

The SC approximation of the Green's function in two
dimensions is\cite{Gutzwiller} 
\begin{eqnarray}
\label{semigreen}
G^{SC}(\mathbf{q},\mathbf{q'},k_F)&=&\frac{2\pi}{(2\pi i\hbar)^{3/2}}
\sum_p\left|D_p(\mathbf{q},\mathbf{q'},k_F)\right|^{1/2}\\
&&\times \exp\left[\frac{i}{\hbar}S_p(\mathbf{q},\mathbf{q'},k_F)
-\frac{i\pi}{2}\mu_p\right],
\nonumber
\end{eqnarray}
where the density of trajectories $D_p$\cite{Gutzwiller} is taken to
vary slowly in comparison with the phase from the action $S_p$ and
$\mu_p$ is called the Maslov index.\cite{Littlejohn,Gutzwiller}

To calculate the $S$-matrix, we start by letting the Hamiltonian act
on the wave function $\psi$ and expand the expression to
\begin{equation}
H\psi=-\frac{\hbar^2}{2m^*}\nabla^2\psi-
\frac{i\hbar e}{2m^*}\nabla\cdot(\mathbf{A}\psi)-
\frac{i\hbar e}{2m^*}\mathbf{A}\cdot\nabla\psi+
\frac{e^2\mathbf{A}^2}{2m^*}\psi.
\end{equation}

The divergence theorem states
\begin{equation}
\int_S d\mathbf{q'}^2\; \nabla\cdot\mathbf{F}=
-\oint_C dl'\;\mathbf{F}\cdot \hat{\mathbf{n}},
\label{divergence}
\end{equation}
where $C$ is the boundary to the area $S$, and $\hat{\mathbf{n}}'$ is
an inward normal to the boundary $C$.
Let 
\begin{equation}
\mathbf{F}=-\frac{\hbar^2}{2m^*}\phi^*\nabla\psi,
\end{equation}
and evaluate
\begin{multline}
\nabla\cdot\mathbf{F}=-\frac{\hbar^2}{2m^*}\phi^*\nabla^2\psi
-\frac{\hbar^2}{2m^*}\nabla\phi^*\nabla\psi=\\
\phi^* H\psi+\frac{i\hbar e}{2m^*}(\phi^*\nabla\cdot(\mathbf{A}\psi)+
\phi^*\mathbf{A}\cdot\nabla\psi)\\
-\frac{e^2\mathbf{A}^2}{2m^*}\phi^*\psi-
\frac{\hbar^2}{2m^*}\nabla\phi^*\cdot\nabla\psi.
\end{multline}
We also get
\begin{equation}
\mathbf{F}\cdot\hat{\mathbf{n}}'=-\frac{\hbar^2}{2m^*}\phi^*
\frac{\partial\psi}{\partial n'}.
\label{fprojection}
\end{equation}
Equations (\ref{divergence}-\ref{fprojection}) results in an analog to
Green's first identity
\begin{multline}
\int_S d\mathbf{q'}^2\; \left[
\phi^* H\psi+\frac{i\hbar e}{2m^*}(\phi^*\nabla\cdot(\mathbf{A}\psi)+
\phi^*\mathbf{A}\cdot\nabla\psi) \right. \\
\left.
-\frac{e^2\mathbf{A}^2}{2m^*}\phi^*\psi- 
\frac{\hbar^2}{2m^*}\phi^*\nabla\psi \right]
= \\
\oint_C dl'\;
\frac{\hbar^2}{2m^*}\phi^*\frac{\partial\psi}{\partial n'}.
\label{Greenident}
\end{multline} 
We take the complex conjugate of equation (\ref{Greenident}),
interchanging $\phi$ and $\psi$, and
subtracting it from equation (\ref{Greenident}), 
after some manipulations we get 
\begin{multline}
\int_S d\mathbf{q'}^2\;\left[ \phi^* H\psi-\psi H\phi^* 
\vphantom{\frac{i\hbar e}{m}}
 \right.\\ \left.
+\frac{i\hbar e}{m}(\psi\mathbf{A}\cdot\nabla\phi^*+
\phi^*\mathbf{A}\cdot\nabla\psi)
\right]=\\
-\frac{\hbar^2}{2m^*}\oint_C dl'\;\left[
\psi\frac{\partial\phi^*}{\partial n'}-
\phi^*\frac{\partial\psi}{\partial n'} \right].
\label{Greentheorem1}
\end{multline}
We would however like to get rid of the magnetic terms on the left
hand side, we now insert 
\begin{equation}
\mathbf{F}=\frac{e\mathbf{A}}{m}\psi\phi^*
\end{equation}
into the divergence theorem and choose a gauge such that
$\nabla\cdot\mathbf{A}=0$, resulting in
\begin{multline}
\int_S d\mathbf{q'}^2\;\frac{i\hbar e}{m}\left[
\psi\mathbf{A}\cdot\nabla\phi^*+\phi^*\mathbf{A}\cdot\nabla\psi\right]=\\
-\oint_C dl'\;\frac{i\hbar e}{m}\mathbf{A}\cdot\hat{\mathbf{n}}'
\psi\phi^*.
\end{multline}
This allows us to rewrite equation (\ref{Greentheorem1}), while replacing
$\phi^*(\mathbf{q'})=G(\mathbf{q},\mathbf{q'})$ as 
\begin{multline}
\label{Greentheorem2}
\int_S d\mathbf{q'}^2\;\left[ G(\mathbf{q},\mathbf{q'}) H\psi-
\psi HG(\mathbf{q},\mathbf{q'}) \right]
=\\
-\oint_C dl'\;\left[\frac{\hbar^2}{2m^*}\left(
\psi\frac{\partial G(\mathbf{q},\mathbf{q'})}{\partial n'}-
G(\mathbf{q},\mathbf{q'})\frac{\partial\psi}{\partial n'}\right) 
\right. \\ \left.
-\frac{i\hbar e}{m}
\psi G(\mathbf{q},\mathbf{q'}) 
\mathbf{A}\cdot\hat{\mathbf{n}}'
\right],
\end{multline}
a version of Green's theorem.

\begin{figure}
\begin{center}
\includegraphics[width=0.4\textwidth]{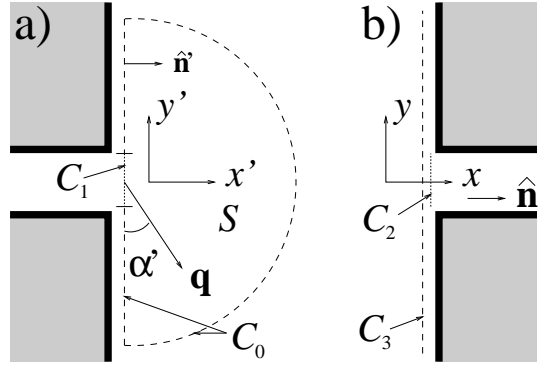}
\end{center}
\caption{a) Geometry of the entrance lead, $S$ is the half circle enclosed
by $C=C_0\cup C_1$, where $C_1$ across the entrance only,
$\hat{\mathbf{n}}'$ is an inward normal to $C$, $\alpha'$ is 
an angle at the entrance, $\mathbf{q}$ is a coordinate and $x'$ and
$y'$ form a local coordinate system. 
b) Geometry of the exit lead, $\hat{\mathbf{n}}$ is into the lead, $x$
and $y$ form a local coordinate system. Path $C_2$ is across the
exit and $C_3$ is along the entire wall.}
\label{leadsfig}
\end{figure}

We now calculate the diffraction from a lead mouth, following
in the steps of Kirchhoff.\cite{Jackson} 
The area $S$ is taken to be the half circle, see figure
\ref{leadsfig}a. The contour $C$ is taken to 
be divided into two parts $C_1$, the entrance 
lead mouth and $C_0$, the rest of $C$. 
We make the requirement that  $\psi=0$ and
$\partial\psi/\partial n'=0$ along the wall, this requirement is
elaborated on later in the text. With zero magnetic field, $G,\psi\sim
r^{-1/2}$, and the integration along the half circle part of $C$ tends to 
zero as the radius $r\rightarrow \infty$.\cite{Schwieters} 
The motivation for letting
$r\rightarrow \infty$ within a finite billiard is that in the
SC theory the wave function along all classical trajectories
is the free space wave function. The total wave function is then the
superposition of all such trajectory wave functions.
In nonzero magnetic field we
are dealing with edge states, localized near the wall. If we assume
that the current is going in the $-y$-direction, then $\psi(\mathbf{q'})$
will vanish  everywhere on the half circle except near the wall in the
lower half-plane $(y'<0)$. On the other hand the Green's function
$G(\mathbf{q},\mathbf{q'})$ is also localized to near the wall but for
$G(y \gg y',y')$, ie. the Green's function at the interior point
$\mathbf{q}$ will not see an excitation at a point $\mathbf{q'}$ that
is far enough downstream the wall. For an interior point $\mathbf{q}$
either $\psi(\mathbf{q'})$ or $G(\mathbf{q},\mathbf{q'})$ will go
(exponentially) fast to zero on the half circle when $r\rightarrow
\infty$.
It is thus enough to integrate over $C_1$ instead of $C$ in equation
(\ref{Greentheorem2}).
Using equations (\ref{Schrodinger}) and (\ref{Green}) in
(\ref{Greentheorem2}) we can
obtain the wave function inside the billiard
\begin{multline}
\psi(\mathbf{q})=\int_{C_1} dy'\;\\
\left[\frac{\hbar^2}{2m^*}\left(
\psi^0(\mathbf{q'})\frac{\partial G(\mathbf{q},\mathbf{q'})}{\partial x'}-
G(\mathbf{q},\mathbf{q'})\frac{\partial\psi^0(\mathbf{q'})}{\partial x'}
\right)
\right.\\ \left.
-\frac{i\hbar e}{m}\psi^0(\mathbf{q'}) G(\mathbf{q},\mathbf{q'})
\mathbf{A}\cdot\hat{\mathbf{n}}'
\right],
\label{Kirchhoff}
\end{multline}
where $dy'=-dl'$.

To calculate transmission amplitudes, we start from the definition of
the current density operator\cite{Messiah}
\begin{equation}
\vec{\mathcal{J}}(\mathbf{q})=\frac{1}{2m^*}\left[\vec{\mathcal P}
\delta(\mathbf{q} - \mathbf{q'})+
\delta(\mathbf{q} - \mathbf{q'}) \vec{\mathcal P}\right],
\end{equation}
where $\vec{\mathcal P}=-i\hbar\nabla+e\mathbf{A}$ is the momentum
operator.
We obtain an operator for the current into the exit lead by
integrating across it
\begin{equation}
{\mathcal J}=\int_{C_2}\! dy\; {\vec{\mathcal J}}(y)\cdot \hat{\mathbf{n}},
\end{equation}
where $C_2$ is across the exit and $y$ the transverse coordinate, see
figure \ref{leadsfig}b.
The eigenmodes $n$ in the entrance lead are taken to be
\begin{equation}
\psi_n^0(x',y')=\frac{1}{\sqrt{v_n}}\xi_n(y')e^{ik_nx'}=
\sqrt{\frac{m}{\hbar k_n}}\xi_n(y')e^{ik_nx'},
\label{eigenmodes}
\end{equation}
where $v_n$ is the velocity in the lead, $k_n$ the k-vector, and $x'$
and $y'$ are  defined in figure \ref{leadsfig}.
The eigenmode wave functions of the exit $\psi_m^0(x,y)$ are defined
analogously. The continuation $\psi_n$ of $\psi_n^0$ into the billiard
is given by equation (\ref{Kirchhoff}).
The scattering matrix element is equal to the matrix element
\begin{multline}
S_{mn}=\langle\psi_m^0|{\mathcal J}|\psi_n\rangle=\\
\int\! dy \left[\frac{-i\hbar}{2m^*}\left(
{\psi_m^0}^*\frac{\partial\psi_n}{\partial x}-
\psi_n\frac{\partial{\psi_m^0}^*}{\partial x} \right)
+\frac{e\mathbf{A}\cdot\hat{\mathbf{x}}}{m}{\psi_m^0}^*\psi_n
\right].
\label{selement}
\end{multline}
By inserting equations (\ref{Kirchhoff}) and (\ref{eigenmodes}) we get
\begin{multline}
S_{mn}=-\frac{i\hbar^3}{4m^2\sqrt{v_mv_n}}
\int dy \int dy' \xi_m^*(y)\xi_n(y') \\
\left[
\frac{\partial^2 G}{\partial x\partial x'}-
\frac{i}{\hbar}\left(mv_n+2e\mathbf{A}(y')\cdot\hat{\mathbf{x}}'\right)
\frac{\partial G}{\partial x} \right.\\\left.
+\frac{i}{\hbar}\left(mv_m+2e\mathbf{A}(y)\cdot\hat{\mathbf{x}}\right)
\frac{\partial G}{\partial x'}\right. \\ \left.
+\frac{1}{\hbar^2}\left(mv_m+2e\mathbf{A}(y)\cdot\hat{\mathbf{x}}\right)
\left(mv_n+2e\mathbf{A}(y')\cdot\hat{\mathbf{x}}'\right)G
\right],
\label{Kirchhoffampl}
\end{multline}
where primed coordinates relate to the entrance and unprimed to the
exit.

Equation (\ref{Kirchhoffampl}) is the transmission amplitude
calculated using the Kirchhoff approximation, which to recall consists
of the following assumptions:
\begin{enumerate}
\item $\psi$ and $\partial\psi/\partial n$
vanish everywhere on $C$ except on the lead mouth. 
\item The values of $\psi$ and $\partial\psi/\partial n$ on the lead
  mouth are equal to the values of the incident wave in the lead.
\end{enumerate}
For example Schwieters et al. \cite{Schwieters} use this
approximation. 
There are however some inconsistencies in the Kirchhoff
approximation.\cite{Jackson}  
It can be shown for the Schr\"odinger equation that if $\psi=0$ and
$\partial \psi / \partial n'=0$ on any finite surface, the the only
solution is $\psi=0$ everywhere. 
This inconsistency can be lifted by choosing Green's functions with
appropriate BC, i.e. either Dirichlet or Neumann. 
We start with the Neumann Green's functions
\begin{equation}
\frac{\partial G_N}{\partial n'}(\mathbf{q},\mathbf{q'})=0 
\text{ for } \mathbf{q'} \text{ on } C.
\label{DirichletGreen}
\end{equation}
If the contour $C$ is an infinite line across the entrance and $S$
all space to the right, the Neumann Green's function can be obtained
by the method of images
\begin{equation}
G_N(\mathbf{q},\mathbf{q'})=
G(\mathbf{q},\mathbf{q'})+
G(\mathbf{q},\mathbf{q''}),
\end{equation}
where $\mathbf{q''}$ is the mirror image of $\mathbf{q'}$, reflected
in $C$.
In the case where $\mathbf{q'}$ lies on $C$, the Neumann Green's
function reduces to
\begin{equation}
G_N(\mathbf{q},\mathbf{q'})=
2G(\mathbf{q},\mathbf{q'}).
\end{equation}
We then get the wave function in the billiard to be
\begin{equation}
\psi=-\int_{C_1} dl'\;
G_N(\mathbf{q},\mathbf{q'})\left[\frac{\hbar^2}{2m^*}
\frac{\partial\psi^0}{\partial n'}
+\frac{i\hbar e}{m}\psi^0\mathbf{A}\cdot\hat{\mathbf{n}}'
\right].
\label{Neumann}
\end{equation}
The coupling to the exit lead should be handled in a way, analog to
the coupling to the entrance lead. The equation for the scattering
matrix element should be symmetric with respect to the direction,
renaming the entrance and exit leads should make no difference. We
therefore need to make the same restriction as on the entrance lead,
i.e. we make a demand on the Green's function on a boundary $C_3$
crossing the exit
\begin{equation}
\frac{\partial G_N}{\partial n}(\mathbf{q},\mathbf{q'})=0 
\text{ for } \mathbf{q} \text{ on } C_3.
\end{equation}
This is again done by the method of images
\begin{equation}
G_{N2}(\mathbf{q},\mathbf{q'})=
G_N(\mathbf{q},\mathbf{q'})+
G_N(\mathbf{q'''},\mathbf{q'}),
\end{equation}
where $\mathbf{q'''}$ is a mirror image of $\mathbf{q}$, reflected in
$C_3$. For $\mathbf{q'}\in C$ and  $\mathbf{q}\in C_3$, we get
\begin{equation}
G_{N2}(\mathbf{q},\mathbf{q'})=
4G(\mathbf{q},\mathbf{q'}).
\end{equation}
Using this Green's function and equation (\ref{Neumann}) in equation
(\ref{selement}), we obtain
\begin{multline}
S_{mn}=-\frac{i\hbar}{m^2\sqrt{v_mv_n}}
\int dy \int dy' \xi_m^*(y)\xi_n(y') \\
\left(mv_m+2e\mathbf{A}(y)\cdot\hat{\mathbf{x}}\right)
\left(mv_n+2e\mathbf{A}(y')\cdot\hat{\mathbf{x}}'\right)G,
\label{Neumannampl}
\end{multline}
which in case of zero magnetic field or if gauge can be chosen such
that $\mathbf{A}\cdot\hat{\mathbf{n}}=0$ and 
$\mathbf{A}\cdot\hat{\mathbf{n}}'=0$, reduces to
\begin{equation}
S_{mn}=-i\hbar\sqrt{v_mv_n}
\int dy \int dy' \xi_m^*(y)\xi_n(y') G,
\label{Neumannampl2}
\end{equation}
which we recognize as the QM expression for the
$S$-matrix. This expression is also most commonly
used for SC calculations. However in QM calculations
the Green's function is derived in a way that includes the effects of
the leads. This expression is exact when using a QM
Green's function.
This is not true for the SC Green's
function which totally ignores the eigenmodes of the leads. Given that
the wave function $\psi$ must be zero on the billiard walls, the choice
of Neumann BC is obviously wrong, since $C$ partially
coincides with the billiard wall. We conclude that this commonly used
expression does not use the correct BC.

The correct BC to use with the SC Green's
function is Dirichlet,
\begin{equation}
G_D(\mathbf{q},\mathbf{q'})=0 
\text{ for } \mathbf{q'} \text{ on } C.
\end{equation}
The Dirichlet Green's function can also be obtained by the method of
images
\begin{equation}
G_D(\mathbf{q},\mathbf{q'})=
G(\mathbf{q},\mathbf{q'})-
G(\mathbf{q},\mathbf{q''}).
\end{equation}
The normal derivative of the Green's function on $C$ can be computed
to
\begin{equation}
\frac{\partial G_D(\mathbf{q},\mathbf{q'})}{\partial n'}=
2\frac{\partial G(\mathbf{q},\mathbf{q'})}{\partial n'},
\end{equation}
for $\mathbf{q'}$ on $C$.
The wave function in the billiard is in this case
\begin{equation}
\psi=\frac{\hbar^2}{2m^*}\int_{C_1} dy'\;
\psi^0\frac{\partial G_D(\mathbf{q},\mathbf{q'})}{\partial x'}.
\label{Dirichlet}
\end{equation}
Following the path of derivation for the Neumann Green's function we
arrive at
\begin{equation}
S_{mn}=-\frac{i\hbar^2}{m\sqrt{k_mk_n}}\int dy \int dy' \xi_m^*(y)\xi_n(y')
\frac{\partial^2 G}{\partial x\partial x'},
\label{Dirichletampl}
\end{equation}
which is our new formula for calculating $S$-matrix elements.

Details on calculating trajectories for the SC Green's
function can be found in reference \cite{TB2}. We need however to take the
derivative of the Green's function too. We will here restrict our self
to zero or weak magnetic field, because the derivatives of the
semi-classical Green's function in magnetic field are very complicated
to compute. In zero magnetic field the action only depends on the
length of the trajectory and it is enough to only look at the ends of
the trajectory to calculate a derivative of the action. In non-zero
magnetic field the action will depend on the total geometry of the
trajectory and  thus also the derivative. In the following
derivation we assume zero magnetic field, and in the end of this
article we show that the resulting formula still yield very good
correspondence to QM, even in finite magnetic field.
The density of trajectories
$D_p$ in equation (\ref{semigreen}) is considered to vary slowly in
comparison to the phase given by the exponential function. The
contribution of one trajectory to the Green's function is considered
to be a plane wave in the vicinity of the trajectory. The
$\partial/\partial x'$-derivative then reduces to taking the derivative
along the trajectory and projecting on the
$\hat{\mathbf{x}}'$-direction, ie. multiplying with $\sin\alpha'_s$,
where the angle  $\alpha'_s$ of the trajectory is defined in figure
\ref{leadsfig}a.
\begin{multline}
\label{semigreenderiv}
\frac{\partial G^{SC}(\mathbf{q},\mathbf{q'},k_F)}{\partial x'}
=\frac{2\pi}{(2\pi i\hbar)^{3/2}}
\sum_p\left|D_p(\mathbf{q},\mathbf{q'},k_F)\right|^{1/2}\\
\times\frac{i}{\hbar}\sin\alpha'_s \frac{S_p(\mathbf{q},\mathbf{q'},k_F)}
{\partial q'_\parallel}
\exp\left[\frac{i}{\hbar}S_p(\mathbf{q},\mathbf{q'},k_F)
-\frac{i\pi}{2}\mu_p\right],
\end{multline}
where $q'_\parallel$ is a coordinate along the trajectory.
The action can be written as
\begin{equation}
S_p=\hbar k_F l_p,
\end{equation}
where $l_p$ is the length of the trajectory. 
We can then calculate a derivative as
\begin{equation}
\frac{\partial S_p}{q'_\parallel}=-\hbar k_F,
\end{equation}
and the Green's function derivative
\begin{multline}
\label{semigreenderiv2}
\frac{\partial G^{SC}(\mathbf{q},\mathbf{q'},k_F)}{\partial x'}
=\\
\frac{-2\pi ik_F}{(2\pi i\hbar)^{3/2}}
\sum_p\left|D_p(\mathbf{q},\mathbf{q'},k_F)\right|^{1/2}\\
\sin\alpha'_s 
\exp\left[\frac{i}{\hbar}S_p(\mathbf{q},\mathbf{q'},k_F)
-\frac{i\pi}{2}\mu_p\right].
\end{multline}
The above expression for the Green's function derivative and the 
corresponding derivative of $x$ have been used in the following
transmission computations.  
We have, as mentioned above, used equation (\ref{semigreenderiv2})
and the corresponding $x$-derivative as approximations in weak
magnetic fields, we have although inserted the exact expression for
the action $S_p$ into these derivative approximations. 
This results in an exact phase but approximate amplitude for the
contribution of a trajectory. 
The resulting good agreement with QM, as will be seen below, justifies
this approximation.  

\begin{figure*}
\begin{center}
\includegraphics[width=0.9\textwidth]{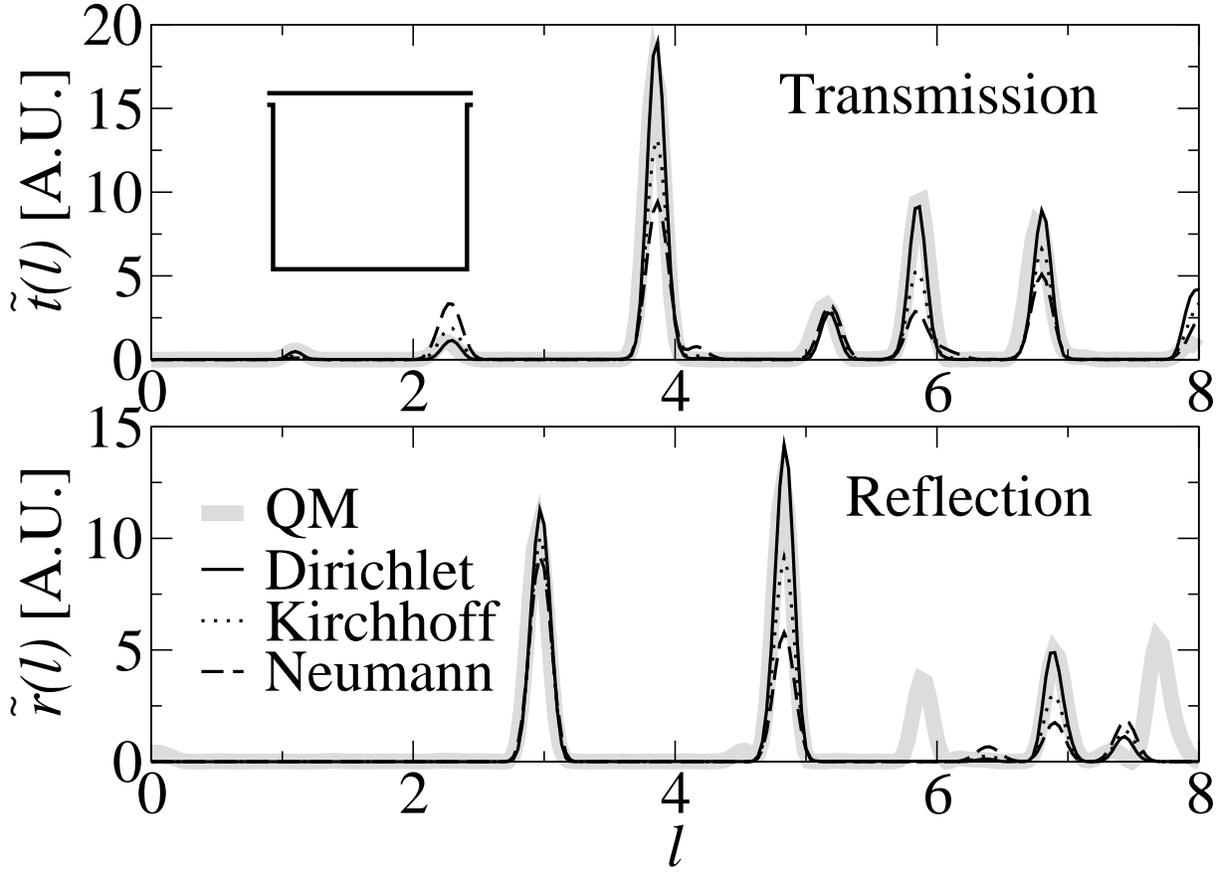}
\end{center}
\caption{The transmission and reflection length spectrum for $S_{11}$
calculated QM and SC using Kirchhoff approximation
[eq. (\ref{Kirchhoffampl})],  Neumann BC [eq. (\ref{Neumannampl})] and
Dirichlet BC [eq. (\ref{Dirichletampl})] for a rectangle, see inset, in
zero magnetic field} 
\label{resultsfig}
\end{figure*}

\begin{figure*}
\begin{center}
\includegraphics[width=0.9\textwidth]{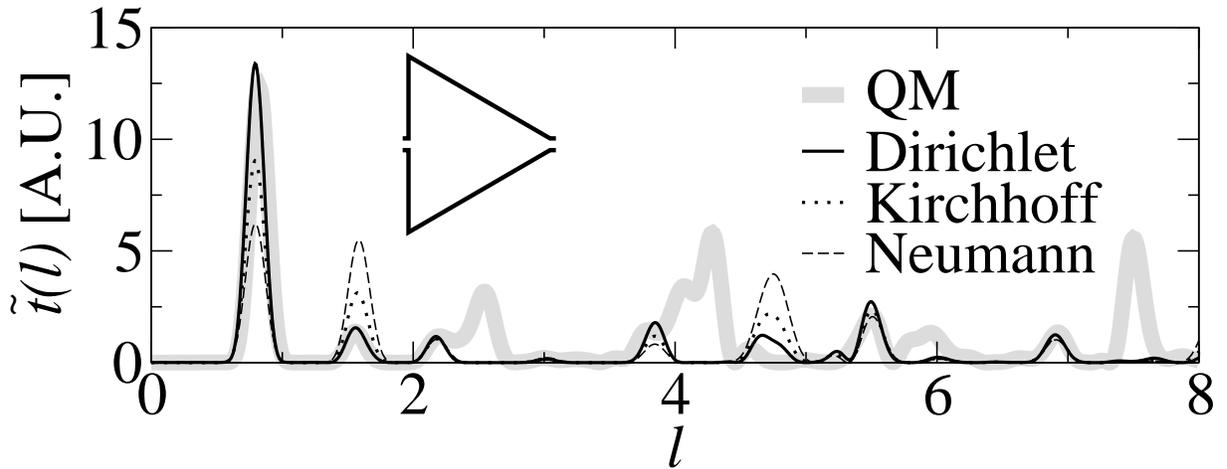}
\end{center}
\caption{The transmission length spectrum for $S_{11}$ calculated QM
and SC using Kirchhoff approximation [eq. (\ref{Kirchhoffampl})],
Neumann BC [eq. (\ref{Neumannampl})] and Dirichlet BC
[eq. (\ref{Dirichletampl})] for a triangle, see inset, in a magnetic
field with cyclotron radius $r_c=0.6L$, where $L$ is side of
triangle.}
\label{resultsfig2}
\end{figure*}

Calculations were made on more than ten different billiards
(triangles and rectangles with different aspect ratio and different
lead positions) of which two are included in this article, see insets
in figures \ref{resultsfig} and \ref{resultsfig2}. The computations
have been made both using QM 
for comparison and using SC theory with the three different  
equations (\ref{Kirchhoffampl}), (\ref{Neumannampl}) and
(\ref{Dirichletampl}). Details on QM calculations can
be found in reference \cite{QMcalc}.
The conductance through the two billiards have been analyzed in terms
of their length spectrum of the transmission/reflection amplitude
defined as  
\begin{equation}
\tilde{t}(l)=\int\!dk_F\; t(k_F,B=\hbar k_F/(er_c))e^{-il k_F},
\end{equation}
where $r_c=\hbar k_F/(eB)$ is the cyclotron radius, and $t$ is the
transmission amplitude, ie. an S-matrix element. 
The length spectrum is thus in magnetic field calculated at constant
cyclotron radius, so the geometry of the trajectories stays constant.
The action along a trajectory in a magnetic field can be written as
\begin{equation}
S_p(\mathbf{q},\mathbf{q'},k_F)=\hbar k_F(l_p+A_pr_c^{-1}),
\end{equation}
where $A_p$ is an area
related to the trajectory. With the density of trajectories varying
slowly with respect to $k_F$ we can see that the length spectrum of the
transmission amplitude will be strongly peaked at
$l=l_p+A_pr_c^{-1}$, or in zero magnetic field $l=l_p$ which
makes the name ``length spectrum'' obvious.

The two analyzed billiards included in this paper, see insets in
figures \ref{resultsfig} and \ref{resultsfig2},
are a rectangle, with sides of length $L$ and $1.1L$ in zero magnetic
field and a 
triangle in magnetic field with $r_c=0.6L$, where $L$ is the side of
the triangle. The leads have a width $w=L/15$, and the transmission
and reflection 
amplitudes has been calculated in the regime of one conducting mode in
the leads. There is a good
correspondence between the positions of the peaks in QM and SC
calculations. The peaks that are missing in SC calculations are due to
``ghost''-trajectories, which bounce diffractively against the lead
mouths. These trajectories are not included in this SC
model.\cite{Schwieters,TB2} 
The included billiard geometries have been
specially chosen to avoid that ``ghost''-trajectories coincide in
length with normal trajectories which otherwise makes comparison with
QM computations difficult. 

\begin{figure}
\begin{center}
\includegraphics[width=0.45\textwidth]{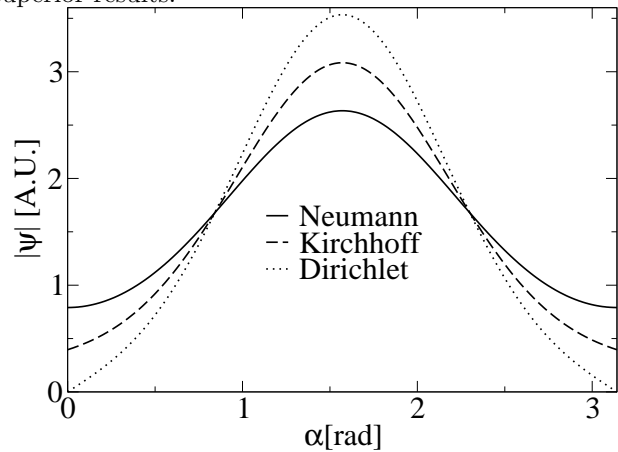}
\end{center}
\caption{The Fraunhofer diffraction pattern from a single lead for
$k_F=1.5\pi/w$, where $w$ is the width of the lead. The plot shows the
amplitude of the wave functions in the far-field as function of
entrance angle $\alpha$, with the three different boundary conditions.}
\label{diffractionfig}
\end{figure}

There is a very good correspondence between SC calculations using
Dirichlet BC [eq. (\ref{Dirichletampl})] and the QM calculations. The
Neumann BC [eq. (\ref{Neumannampl})] still gets the peaks in the right
positions but can not reproduce the correct amplitude. The Kirchhoff
approximation [eq. (\ref{Kirchhoffampl})] can
be seen as a mean value between Neumann and Dirichlet BC and
produce peak amplitudes between the Neumann and Dirichlet
amplitudes. It is interesting to note that even in the case of the
triangle where the geometry near one of the leads differs from the
case in the derivation and we are using an only approximate expression
for the Green's function derivative, the Dirichlet
BC still gives superior results (there is no derivative in the
expression for the Neumann BC transmission amplitude, so it is not
affected by the approximation). 
The actual difference between the usage of the three
different BC is, at least in zero magnetic field, in the diffraction
from the leads. The different BC yield different diffraction patterns,
see figure \ref{diffractionfig}.  

To conclude, we have studied the effect of the choice of boundary
conditions at the lead mouths.
We find that in contrast to prior
practise of using an expression from QM that agrees with Neumann BC,
or using the Kirchhoff approximation, one should use Dirichlet BC.
This is not in disagreement with QM because the QM Green's function
differs from its SC counterpart in that it includes the effect of the
eigenmodes in the leads, while the SC Green's function does not. The
SC Green's function is only an approximation of the QM Green's
function and because of its ignorance of the wave function outside of
the billiard it should be treated differently from the QM Green's function.
We therefore propose a new expression for the scattering matrix,
equation (\ref{Dirichletampl}). By comparison to QM calculations, we
show this expression to yield superior results.

\begin{acknowledgments} Financial support from the National
Graduate School in Scientific Computing is acknowledged. I thank
I. V. Zozoulenko for valuable discussions.
\end{acknowledgments}

\end{document}